# Atomic visualization of copper oxide structure in infinite-layer cuprate SrCuO$_2$

Yong Zhong[1], Sha Han[1], Yang Wang[1], Zhiling Luo[1], Ding Zhang[1,2], Lili Wang[1,2], Wei Li[1,2], Ke He[1,2], Can-Li Song[1,2,*], Xu-Cun Ma[1,2,*], and Qi-Kun Xue[1,2,*]

[1] *State Key Laboratory of Low-Dimensional Quantum Physics, Department of Physics, Tsinghua University, Beijing 100084, China*

[2] *Collaborative Innovation Center of Quantum Matter, Beijing 100084, China*

We report the atomic-scale structure of epitaxial films of parent infinite-layer compound SrCuO$_2$ prepared on SrTiO$_3$ by molecular beam epitaxy. *In-situ* scanning tunneling microscopy study reveals a stoichiometric copper oxide (CuO$_2$)-terminated surface featured by 2 × 2 reconstruction, caused primarily by structural distortions of four adjacent CuO$_2$ plaquettes. Furthermore, the subsurface Sr atoms have been rarely discernible, showing intra-unit-cell rotational symmetry breaking. These observations can be reasonably modelled by a periodic up-down buckling of oxygen ions on the CuO$_2$ plane. Further post-annealing leads to the removal of surface oxygens and an incommensurate stripe phase. Our study provides indispensable structural information to help understand the exotic properties of cuprate superconductors.

*\* To whom correspondence should be addressed. Email: clsong07@mail.tsinghua.edu.cn, xucunma@mail.tsinghua.edu.cn, qkxue@mail.tsinghua.edu.cn*

The underlying mechanism of high temperature ($T_c$) superconductivity in cuprates remains one of the most mysterious issues in condensed matter physics [1, 2], although more than thirty years have elapsed since the discovery. This is in part due to the complex structures of cuprate superconductors, in which the copper oxide ($CuO_2$) planes are sandwiched between other ionic layers. The ionic layers often serve as sources of electrons or holes (charge reservoirs) for $CuO_2$ planes, thereby doping them and inducing superconductivity there. To unveil the high-$T_c$ superconductivity in cuprates, many experimental techniques including surface-sensitive scanning tunneling microscopy (STM) and angle-resolved photoemission spectroscopy have widely been applied and demonstrated their power to reveal cuprates' intriguing physical properties [3, 4]. These include the generally accepted $d$-wave pairing symmetry, pseudogap, electronic inhomogeneity, a broad class of charge/spin orders and nematicity [5-9]. Unfortunately, however, these measurements are mostly taken on charge reservoir layers rather than the $CuO_2$ plane, because the latter is hardly accessible to surface sensitive techniques. This leaves an open question whether the measurements can actually reflect the properties of superconducting $CuO_2$ planes. Such concerns are demonstrably justified by a few experimental studies performed directly on the $CuO_2$ plane [10-13], which reveal contrasting features as those previously reported in cuprates. In order to clarify the novel phenomena and pairing mechanism of cuprate superconductors, preparation and exploration of the $CuO_2$ plane are desired but challenging in experiments.

Among cuprates, the infinite layer (IL) $ACuO_2$ ($A$ = Ca, Sr, Ba) compounds have the simplest crystal structure and a stacking of alternative $CuO_2$ and Sr planes. The partial substitution of Sr by La introduces electrons and consequently leads to superconducting $Sr_{1-x}La_xCuO_2$ ($x$ = 0.05 ~ 0.15) with a maximum transition temperature of $T_c$ = 43 K [14, 15]. This value can be further enhanced to 80 K by constructing the $BaCuO_2/CaCuO_2$ superlattice [16], and to 110 K by isovalent substitution in hole-doped $(Sr_{1-x}Ca_x)_{1-y}CuO_2$ ($y$ = 0.1) [17]. More significantly, a nodeless superconducting gap, in contradiction to $d$-wave nodal scenario, is suggested by both bulk [18] and surface [19, 20] sensitive measurements in optimal doped IL $Sr_{0.9}La_{0.1}CuO_2$. Besides, the IL cuprates commonly contain no apical oxygen and are readily terminated with $CuO_2$ plane, the key building block of cuprate superconductors. All these characteristics render IL cuprates ideal prototype systems to directly exploit the key $CuO_2$ plane and resolve the long-standing puzzles in cuprates. For example, a comparison study of IL cuprates with the widely studied hole-doped cuprates will certainly help understand the roles played by apical oxygens in the high-$T_c$ superconductivity of hole-doped cuprates. However, a systematic atomic-scale characterization of IL cuprates is currently lacking.

Despite granular bulk counterparts by high-pressure solid-state reaction [14, 15], epitaxial films of IL $Sr_{1-x}La_xCuO_2$ have recently been synthesized by state-of-the-art molecular beam epitaxy (MBE) [21, 22]. In this study, we mainly focus on the parent IL $SrCuO_2$ films prepared on $SrTiO_3$ (001) substrates, and detect their atomic-scale structure by *in-situ* STM. All the experiments were carried out in an ultrahigh vacuum (UHV) system (Unisoku) combining both MBE and STM capabilities. The base pressure of both chambers is better than $1.0 \times 10^{-10}$ Torr. Nb-doped $SrTiO_3$ (0.05 wt%) substrates were annealed at 1200°C for 30 minutes to obtain $TiO_2$-terminated surface, and kept at 650°C during the subsequent film growth. IL $SrCuO_2$ thin films were prepared by co-evaporating high-purity Cu (99.9999%) and Sr (99.95%) sources from their respective effusion cells under an ozone pressure of $1.5 \times 10^{-5}$ Torr. Once the film growth was finished, the $SrCuO_2$ epitaxial films were *in-situ* transferred into STM chamber for data collection at 4.2 K. Polycrystalline PtIr tips were conditioned by electron beam bombardments and calibrated on Ag/Si (111) films prior to all STM measurements in a constant-current (10 pA) mode.

Figure 1(a) shows a schematic view of IL $SrCuO_2$ epitaxial films prepared on $TiO_2$-terminated $SrTiO_3$ (001) substrates. IL $SrCuO_2$ is generally described as an oxygen deficient perovskite, with missing oxygen atoms in the Sr layers. The in-plane lattice parameter is $a = b = 3.94$ Å, while the out-of-plane one is $c = 3.43$ Å. For the epitaxial growth of $ACuO_2$, previous studies have revealed thickness-dependent phase transition from IL tetragonal to chain-type $SrCuO_2$ as the films are thinned down to 4 unit cells (UC) [22-24]. Distinct from IL $SrCuO_2$ [Fig. 1(a)], half of $O^{2-}$ ions are moved to the $Sr^{2+}$ plane in the chain-type structure, which leads to a stacking of alternative CuO and SrO planes and an increase of ~ 0.5 Å in the *c*-axis lattice parameter ($c = 3.91$ Å). Notably that the undoped IL $SrCuO_2$ behaves as a Mott insulator, which poses great challenges for STM characterization. To overcome this difficulty, ultrathin $SrCuO_2$ films with a thickness of 6 UC are prepared on the as-cleaned $SrTiO_3$ (001) substrates. Shown in Fig. 1(b) is a typical STM topography of $SrCuO_2$ film, which grows in a layer-by-layer manner. To confirm that the epitaxial films are indeed IL $SrCuO_2$ rather than chain-type one, a thicker film of 20 UC was prepared and characterized by X-ray diffraction (XRD). As demonstrated in Fig. 1(c), except for XRD signals from SrTiO$_3$ substrate, the (001) and (002) XRD peaks at 25.7° and 51.9°, characteristic of IL SrCuO$_2$, are evident. Calculated from the peak positions, we obtain a lattice parameter of $c = 3.48$ Å, as anticipated for IL SrCuO$_2$. Such a value is further confirmed by taking the line profile across a step edge in Fig. 1(d). We estimate the step height to be about 3.5 Å, which is in good agreement with the XRD measurements. Notably, the measured $c$-axis lattice constant appears slightly larger than that of IL (3.43 Å) bulk SrCuO$_2$, suggesting that the epitaxial films are compressively strained.

Shown in Fig. 2(a) is a magnified STM image acquired on the flat terrace of IL SrCuO$_2$ films. Interestingly, checkboard-like square lattice with a spacing of ~ 7.9 Å, at twice the bulk periodicity, is revealed and indicative of the formation of 2 × 2 reconstructed surface. This suggests that the SrCuO$_2$ films are terminated with CuO$_2$ [22], which provide us an unprecedented opportunity to characterize directly its structure on the atomic scale. Otherwise, the pristine 1 × 1 Sr lattice should have been observed [25]. The doubly-periodic surface reconstruction is more clearly seen from the Fast Fourier transform (FFT) image as displayed in Fig. 2(b). This always holds true irrespective of where we image the atomic-scale structures. Based on the Bragg spots (black circles), we calculate the in-plane lattice constants from all the FFT images we obtained, leading to an averaged value of 3.93 ± 0.03 Å. This value justly lies between the in-plane lattice parameter of SrTiO$_3$ (3.905 Å) substrate and bulk SrCuO$_2$ (3.94 Å), consistent with the compressive SrCuO$_2$ films. Additionally, the four peaks marked by orange circles correspond to the 2 × 2 surface reconstruction, while the two orthogonal 2 × 4 and 4 × 2 superstructures in Fig. 2(a), most probably caused by missing surface Cu atoms, occasionally appear on the CuO$_2$ plane and contribute to the cyan-circled peaks in Fig. 2(b).

For a better understanding of the CuO$_2$-(2 × 2) surface reconstruction on the atomic scale, we acquire a series of zoom-in STM images at varying occupied states, as shown in Fig. 2(c). Obviously, the STM images acquired at smaller sample biases give finer structures of the 2 × 2 reconstruction, which are more clearly revealed from the line profiles in Fig. 2(d). Every unit cell is found to consist of four bright spots, which preferentially shrink together to form the tetramer-like patterns. In IL SrCuO$_2$, the nearest-neighboring O-O distance is approximately 2.7 Å, significantly smaller than the spacing between adjacent bright spots. We thus ascribe the bright spots as Cu cations. This is consistent with the local density approximation (LDA) calculations [26], which shows Cu $3d_{x^2-y^2}$ orbital-dominant electronic states near -4 eV. Our finding represents the first real-space observation that the 2 × 2 reconstructed surface of $CuO_2$ plane is primarily caused by preferential structural distortions of four adjacent $CuO_2$ plaquettes. We confirm this by registering the lateral position of every Cu atoms (short vertical lines) along the [010] azimuth in the bottom panel of Fig. 2(d). The alternative short (~ 3.7 Å) and long (~ 4.2 Å) variations of the interatomic spacing are revealed. This means that the Cu atoms have moved from their equilibrium position by 0.125 Å (~ 3.2%) along both the *a* and *b* axes, leading to the *a* and *b*-orientated dark valleys in Figs. 2(a) and 2(c). The height difference between the Cu sites and dark valleys can be as high as 15 pm.

Another prominent feature we observe here is that the valleys along the *a* axis appear darker than those along the *b* axis (c.f. the blue and black curves in the top panel of Fig. 2(d)). This suggests that except for the structural distortion-induced 2 × 2 reconstruction, the $CuO_2$ plane additionally experiences a rotational symmetry breaking from $C_4$ to $C_2$. Such a claim receives more experimental evidences from occasionally imaging the subsurface Sr cations. As shown in Fig. 3a, we visualize the subsurface Sr atoms at a smaller sample bias of -3.6 V, although the identical STM tip sees the top Cu atoms at -4.0 V as usual (c.f. the inset of Fig. 3(a) and middle panel of Fig. 2(c)). For easy viewing, the cross symbol denotes the identical position in both STM images at different sample voltages. One can immediately note that the dark holes in the inset of Fig. 3(a) correspond to the brightest spots in Fig. 3(a). Knowing that the dark holes are positioned at the hollow sites of top Cu lattice, the bright spots in Fig. 3(a) must the subsurface Sr atoms. Remarkably, although Sr atoms (dubbed as Sr2) between the brightest spots (Sr1) can be resolved along the *a* axis, the equivalent Sr3 atoms along the *b* axis is little identifiable [Fig. 3(b)]. This supports an intra-unit-cell symmetry breaking of the $CuO_2$ plane. Such a finding constitutes one of the major findings in this study and is of significant importance, because it might link intrinsically to the extensively studied symmetry breaking states and nematicity in cuprates and provides atomic scale insight into their mechanism [27, 28].

Our findings appear irreconcilable with the recently proposed oxygen vacancy model for 2 × 2 surface reconstruction of $SrCuO_2$ [22], in which half oxygen is assumed to be dismissed in every unit cell to avoid the polar catastrophe due to the alternating stacking of positively charged $Sr^{2+}$ and negatively charged $CuO_2^{2-}$ planes [29]. In the atomically resolved STM images [Figs. 2(c) and 3(a)], we find little signature of point vacancies, suggesting high stoichiometry of as-grown $SrCuO_2$ films. Moreover, the supposed oxygen vacancies wholly break the $C_4$ symmetry of host lattice to $C_1$, which is contradictive to our observation of $C_2$-symmetric STM features. Instead, we propose here a periodic up-down buckling of oxygen ions on the $CuO_2$ plane, as illustrated in Fig. 3(c). This scenario sounds more reasonable since the buckling of oxygen ions has been previously observed in several cuprate compounds [30, 31] as well as the isostructural $SrFeO_2$ compound [32]. In this model, no oxygen vacancy is involved and the buckling of $CuO_2$ plane by displacing the oxygen ions in an up-down manner might modify the charge distribution and thus stabilize the terminated $CuO_2$ plane [33]. As shown in Fig. 3(c), the plus and minus signs (+ and −) mark the oxygen ions shifted upward and downward relative to the $CuO_2$ plane, respectively, while the size of green spheres signifies the magnitude of displacement. As a result, the oxygen ions with a larger buckling tend to shrink the nearest-neighboring two Cu atoms together (see the arrows in Fig. 3(c)), eventually leading to the appearance of Cu tetramers. This naturally interprets the alternative short and long interatomic distances in Fig. 3(d) and formation of $2 \times 2$ superstructure.

More significantly, a close-up inspection of the model in Fig. 3(c) reveals directly the symmetry breaking from $C_4$ to $C_2$ of the $CuO_2$ plane, as the directions of oxygen buckling are concerned. Given the variations of chemical environment, there exist four categories of Sr atoms (denoted as Sr1, Sr2, Sr3 and Sr4). Different from Sr1 and Sr2 bound to the oxygens with small downward buckling, the oxygens adjacent to Sr3 and Sr4 exhibit large downward buckling. As such, the stronger interactions between the buckled oxygens and Sr3/Sr4 might pull significantly the corresponding Sr atoms downwards, making them invisible for STM. This matches excellently with our observations in Figs. 3(a) and 3(b), where only halves of subsurface Sr atoms are visualized. Furthermore, as compared to Sr2, the Sr1 atoms are located in the hollow sites of four adjacent Cu tetramers. One can naively suppose that they should more easily be seen and appear as the brightest atoms, consistent with our experiments. Although this model is subject to further experimental and theoretical confirmation, our findings of Cu tetramers and intra-unit-cell symmetry breaking place strong constraints on any atomic models of the $CuO_2$ plane in parent IL $SrCuO_2$ cuprates.

Finally we argue that a model with stoichiometric $CuO_2$, rather than the $CuO_{1.5}$ oxygen vacancy scenario, is further supported by intentionally removing the surface oxygens. Figure 4(a) shows a typical STM topography of $SrCuO_2$ films after being annealed at 500 °C in UHV. Instead of the $2 \times 2$ superstructure, one-dimensional and uniformly spaced (~ 2.7 nm) stripes are always observed and orientated along either *a* or *b* axis. Such stripe structures occur as well on IL $SrCuO_2$ epitaxial films prepared at an elevated substrate temperature of 720 °C, which equivalently corresponds to oxygen-deficient environment. This hints that the 2 × 2 superstructure is not related to oxygen vacancies. Figure 4(b) displays a zoom-in STM image of the striped surface. The bright spots along the stripes are spaced at ~ 1.6 nm (4$a$) [Fig. 4(c)], suggestive of a larger commensurate superstructure. Although the 2.7 nm spacing between adjacent stripes coincides closely with the periodicity of the well-documented supermodulations in $Bi_2Sr_2CaCu_2O_{8+\delta}$ [34], it is worth noticing that the latter supermodulations are orientated along one of the diagonal directions. Further studies are needed to understand this discrepancy as well as their formation mechanism.

In summary, we have clarified the atomic-scale structure of parent IL $SrCuO_2$ cuprate films in real-space by using cryogenic scanning tunneling microscopy. This enables the building of a periodic up-down buckling model of oxygen ions on $CuO_2$ plane that accounts well for the doubly-periodic surface reconstruction and intra-unit-cell rotational symmetry breaking. Knowing that the superconducting IL $Sr_{1-x}La_xCuO_2$ involves no surface reconstruction and develops from the parent $SrCuO_2$ compounds [15, 22], our study has stepped towards understanding the superconductivity of IL cuprates on the atomic scale.

**Acknowledgements**: We acknowledge Andrey Oreshkin (MSU) for the helpful discussion. This work was financially supported by the Ministry of Science and Technology of China (Grants No. 2017YFA0304600, 2015CB921001, 2016YFA0301004), the National Natural Science Foundation of China (Grants No. 11427903, 11504196, 11634007, 11774192), and in part by the Beijing Advanced Innovation Center for Future Chip (ICFC). C. L. S. acknowledges support from the National Thousand-Young-Talents Program and Tsinghua University Initiative Scientific Research Program.

**Figure captions**

**FIG. 1** (color online). (a) Schematic sketch of IL $SrCuO_2$ epitaxial films grown on $TiO_2$-terminated $SrTiO_3$ (001) substrate. (b) STM topography (100 nm × 100 nm, $V$ = -5 V) of 6 UC $SrCuO_2$ film. Inset shows the top view of $SrCuO_2$ involving four $CuO_2$ plaquettes. (c) XRD pattern of a 20 UC $SrCuO_2$ epitaxial film. Peaks characteristic of the $SrCuO_2$ film and $SrTiO_3$ substrate are indicated. (d) Line profile taken along the dashed line in (b), showing a uniform step height of 3.5 Å in the epitaxial $SrCuO_2$ film.

(b)inset: the color of Cu atom should be modified to the same color as in (a)

**FIG. 2** (color online). (a) Enlarged STM topography (20 nm × 20 nm, $V$ = -4.2 V) acquired on the flat $SrCuO_2$ terrace, showing a 2 × 2 checkboard structure. The black square marks the surface unit cell of 2 × 2 superstructure throughout. (b) Fourier transform image of Fig. 2(a). Peaks marked by colored circles correspond to the primitive 1 × 1 Cu lattice (black), 2 × 2 superstructure (orange), and orthogonal 2 × 4 and 4 × 2 superstructures (cyan), respectively. (c) Atomic resolution images (4.6 nm × 4.6 nm) at various sample biases as indicated, acquired in the dashed square marked region of (a) . (d) Line profiles along the $a$ (black curve) and $b$ (blue curve) axes. Short vertical lines at the bottom panel mark the positions of every Cu atom along the $b$ axis, showing the short (~ 3.7 Å) and long (~ 4.1 Å) alternative variations of the interatomic distances.

**FIG. 3** (color online). (a) STM image (5.6 nm × 5.6 nm) of subsurface Sr atoms taken at a small sample bias of -3.6 V, with only Sr1 and Sr2 atoms visible. Inset shows the STM image in the same field of view at a higher bias of -4.0 V. (b) Line profiles along the $a$ and $b$ axes. (c) Up-down oxygen buckling model. The plus (+) and minus (−) signs mark the oxygen ions up- and down-buckled with respect to the $CuO_2$ plane, respectively, with the size of green spheres denoting the buckling strength. The shaded plum spheres correspond to the Cu positions before clustering.

**FIG. 4** (color online). (a) STM topography (50 nm × 50 nm, $V$ = -5 V) of $SrCuO_2$ films after post-annealing at 500 ºC for 30 minutes, showing an incommensurate and 2.7 nm-spaced stripe phase. (b) Zoom-in image (50 nm × 10 nm, $V$ = -4 V) of the striped phase. (c) Line profiles showing a commensurate superstructure of 4$a$ along the stripes in (b).

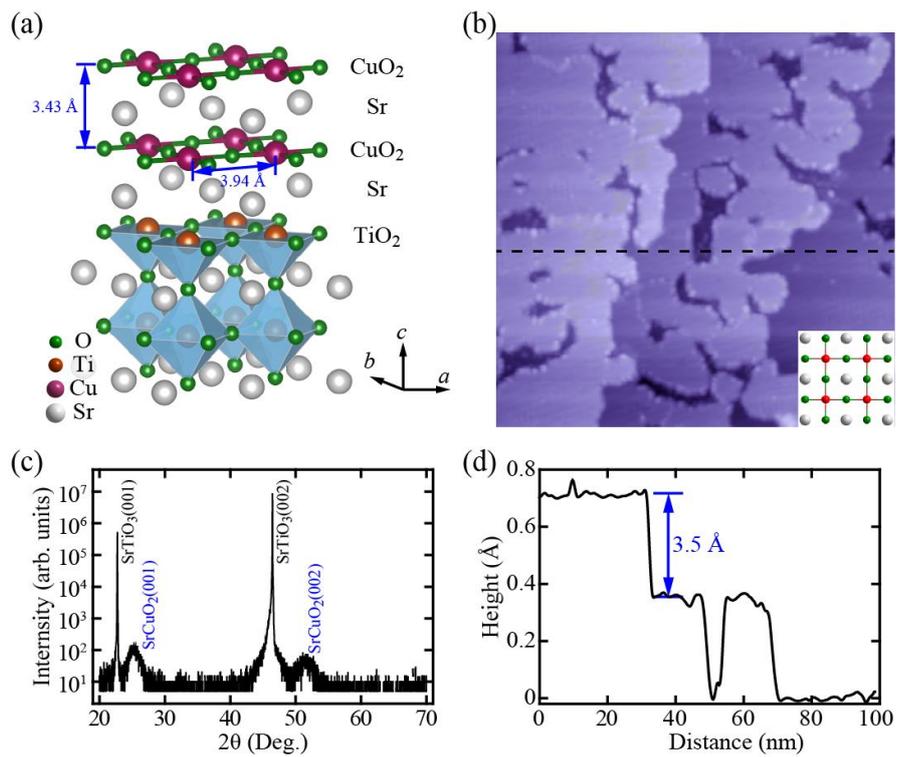

**Figure 1**

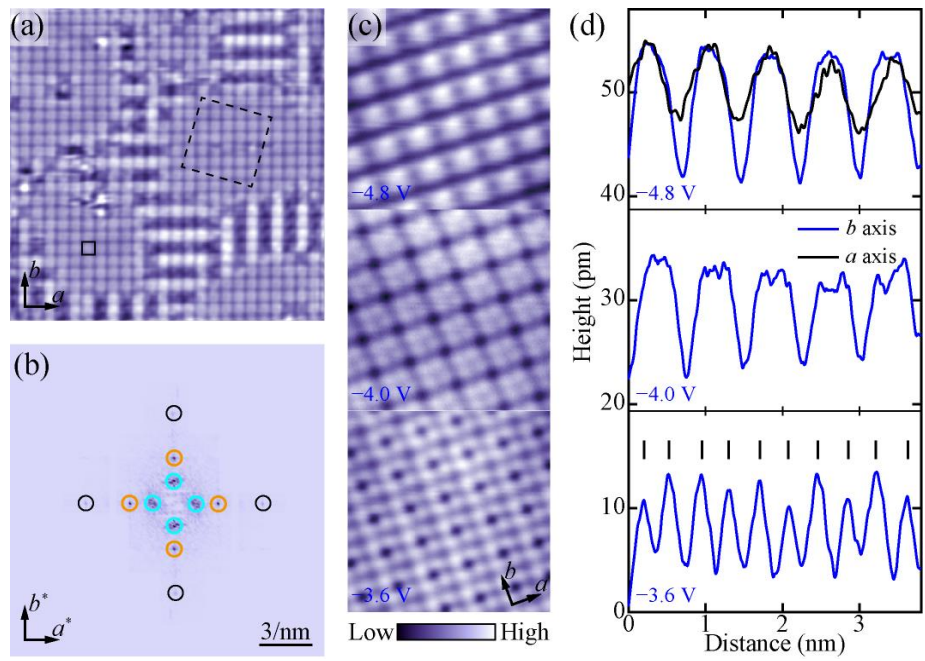

**Figure 2**

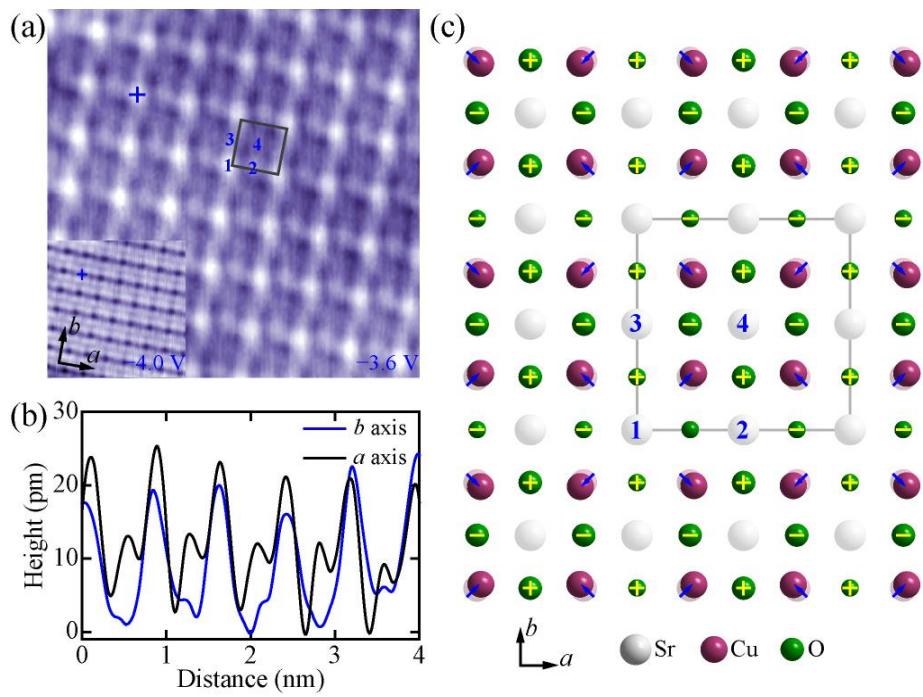

**Figure 3**

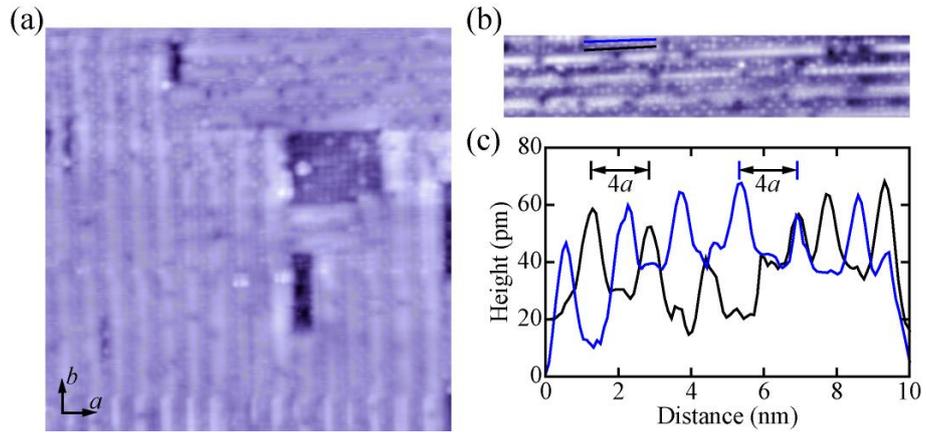

**Figure 4**